\documentclass[fleqn,usenatbib]{mnras}

\usepackage{newtxtext,newtxmath}

\usepackage[T1]{fontenc}

\DeclareRobustCommand{\VAN}[3]{#2}
\let\VANthebibliography\thebibliography
\def\thebibliography{\DeclareRobustCommand{\VAN}[3]{##3}\VANthebibliography}

\usepackage{graphicx}	
\usepackage{amsmath}	
\usepackage{bm}
\usepackage{blindtext}
\usepackage[utf8]{inputenc}

\newcommand{\mearth}{${\,{\rm M}_{\oplus}}$}
\newcommand{\rearth}{${\,{\rm R}_{\oplus}}$}

\title[Detecting Nightside City Lights]{The Detectability of Nightside City Lights on Exoplanets}

\author[T. G. Beatty]{
Thomas G. Beatty$^{1}$\thanks{email: tgbeatty@arizona.edu}
\\
$^{1}$Department of Astronomy and Steward Observatory, University of Arizona, Tucson, AZ 85721
}

\date{Accepted XXX. Received YYY; in original form ZZZ}

\pubyear{2022}

\begin{document}
\label{firstpage}
\pagerange{\pageref{firstpage}--\pageref{lastpage}}
\maketitle

\begin{abstract}
Next-generation missions designed to detect biosignatures on exoplanets will also be capable of placing constraints on technosignatures (evidence for technological life) on these same worlds. Here, I estimate the detectability of nightside city lights on habitable, Earth-like, exoplanets around nearby stars using direct-imaging observations from the proposed LUVOIR and HabEx observatories, assuming these lights come from high-pressure sodium lamps. I consider how the detectability scales with urbanization fraction: from Earth's value of 0.05\%, up to the limiting case of an ecumenopolis -- or planet-wide city. Though an Earth analog would not be detectable by LUVOIR or HabEx, planets around M-dwarfs close to the Sun would show detectable signals at $3\,\sigma$ from city lights, using 300 hours of observing time, for urbanization levels of 0.4\% to 3\%, while city lights on planets around nearby Sun-like stars would be detectable at urbanization levels of $\gtrsim10\%$. The known planet Proxima b is a particularly compelling target for LUVOIR A observations, which would be able to detect city lights twelve times that of Earth in 300 hours, an urbanization level that is expected to occur on Earth around the mid-22nd-century. An ecumenopolis, or planet-wide city, would be detectable around roughly 30 to 50 nearby stars by both LUVOIR and HabEx, and a survey of these systems would place a $1\,\sigma$ upper limit of $\lesssim2\%$ to $\lesssim4\%$, and a $3\,\sigma$ upper limit $\lesssim10\%$ to $\lesssim15\%$, on the frequency of ecumenopolis planets in the Solar neighborhood assuming no detections.
\end{abstract}




\section{Introduction}

The search for biosignatures on potentially habitable exoplanets has driven much of the recent planning in the exoplanet community for the next generation of space telescopes. In particular, the National Academies' Exoplanet Science Strategy and Astrobiology Strategy reports both recommended the development of a large, space-based, direct-imaging observatory that would be capable of directly detecting emission from habitable Earth-like exoplanets. At the moment, the two primary observatory architectures under consideration are the Large UV/Optical/IR Surveyor \citep[LUVOIR,][]{luvoirfinal} and the Habitable Exoplanet Observatory \citep[HabEx,][]{habexfinal}. Either would provide a dramatic increase in our ability to characterize exoplanet atmospheres and search for biosignatures.

\begin{figure*}[t]
\vskip 0.00in
\includegraphics[width=1\linewidth,clip]{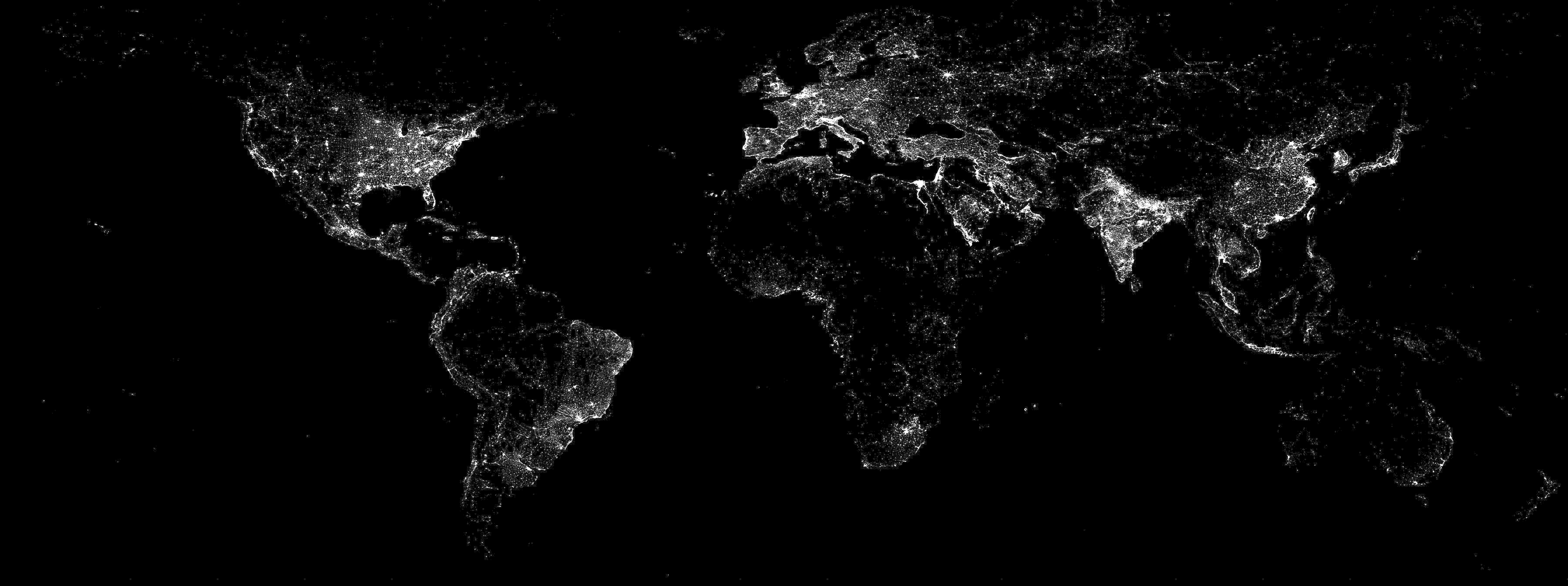}
\vskip -0.0in
\caption{The nightside of Earth shows significant emission from city lights in the optical. This is a composite, cloud-free, image of Earth's city lights compiled using Day/Night Band observations taken using the Visible Infrared Imaging Radiometer Suite instrument on the Soumi National Polar-orbiting Partnership satellite \citep{blackmarbleATBD}. Searching for the emission from city lights is a compelling technosignature because it requires very little extrapolation from current conditions on Earth, should be relatively long-lived presuming an urbanized civilization, and offers a very distinct spectroscopic signature that is difficult to cause via natural processes. This places the emission from nightside city lights high on the list of potential technosignatures to consider \citep{sheikh2020}.}
\label{blackmarble}
\end{figure*}

As our ability to characterize exoplanets improves, we also move closer to the direct detection of technosignatures on exoplanets. 
Similar in concept to biosignatures, technosignatures are observable signals which indicate the presence of technological life on an exoplanet or in a star system \citep{tarter2007,setivocab}. An ideal technosignature would be a long-lived, unambiguous, observable property of another civilization that (for simplicity) requires relatively little extrapolation from conditions on Earth \citep{sheikh2020}.  

For the last several decades the main methods of searching for technosignatures have been radio \citep[e.g.,][]{breaklisten} or optical \citep[e.g.,][]{howard2004} monitoring projects to look for signals actively being broadcast by a nearby technological civilization. The lack of a technosignature detection from these projects implies that the frequency of other technological life in the Solar neighborhood is low, that other technological life is not strongly broadcasting in the radio or optical, that our detection sensitivities are low, or some combination of these three.

The development over the last fifteen years of the facilities and techniques necessary to precisely characterize the atmospheres of exoplanets gives us a new way to search for technosignatures, and to do so in a way that does not rely on active broadcasts, by directly examining potentially habitable planets themselves. While this approach will likely result in narrower searches (in terms of number of targets) than the radio and optical broadcast surveys, directly searching for technosignatures on exoplanets is sensitive to a much wider range of possible phenomena \citep[e.g.,][]{kipping2016,wright2016}. Assuming such a direct search finds no significant detections, this also likely means that tighter constraints could be placed on the last three terms in the Drake Equation \citep[which encapsulate the frequency and lifetime of intelligent civilizations,][]{drakeeq}, than for a typical radio or optical broadcast survey.

Searching for technosignatures on exoplanets also has the virtue of being able to piggy-back on the observations necessary to detect biosignatures on these same planets. For example, the deep spectroscopic observations needed to successfully detect biosignatures on an Earth-analog exoplanet using either LUVOIR and HabEx \citep[e.g.,][]{wang2018} will also be capable of detecting -- or at least meaningfully constraining -- certain technosignatures. In this sense many of these observations will be ``free'' in terms of observing time, and the expected signals from likely technosignatures should be looked for in these future data sets.

Already, significant thought has been put into what some of these technosignatures might look like. \cite{loeb2017} considered the technosignature on the surface of an Earth-like exoplanet caused by large-scale use of silicon-based photovoltaic arrays, which would show a significantly different reflection spectrum than vegetation. \cite{berdyugina2019} considered the more general problem of identifying artificial surface features on Proxima b, which might be possible using next-generation space telescopes. \cite{exobelt} investigated the observational signatures of dense satellite constellations in orbit about an Earth-like exoplanet, and found that for Earth-analogs with M-dwarf primaries this may be observable in transit data. \cite{vides2019} determined that relatively low powered (10s of kW) laser pulses would be detectable from nearby Sun-like stars in direct imaging observations from the Gemini Planet Imager and the Roman Space Telescope. A bit more grimly, \cite{stevens2016} noted that the increased atmospheric dust and clouds caused by a large nuclear war could be observable on an exoplanet, though it would likely be difficult to distinguish this from the impact of a large asteroid. 

More specifically on possible atmospheric technosignatures, both \cite{lin2014} and \cite{ravi2021} have investigated the detectability of industrial pollution in an Earth-analog's atmosphere. \cite{lin2014} looked at chlorofluorocarbons, since these are extremely unlikely to be generated via biotic or geologic processes, and found concentrations ten times that in Earth's atmosphere would be detectable using 30 to 50 hours of time on the James Webb Space Telescope for Earth-analogs orbiting white dwarfs. \cite{ravi2021} considered pollution from nitrogen dioxide and estimated that the 15m LUVOIR A architecture would be capable of detecting Earth's level of nitrogen dioxide pollution using 400 hours of observing time on a system at 10 pc.

The emission from city lights on the nightsides of exoplanets is another potential technosignature. Conceptually, city lights are a compelling technosignature because it requires very little extrapolation from current conditions on Earth, should be relatively long-lived presuming an urbanized civilization, and offers a very distinct spectroscopic signature that is difficult to cause via natural processes. This places the emission from nightside city lights high on the list of potential technosignatures to consider \citep{sheikh2020}, and the idea of searching for city lights on exoplanets has been suggested before by \cite{schneider2010} and \cite{loeb2012}.

Here, I consider the detectability of nightside city lights on habitable, Earth-like, exoplanets around nearby stars in the optical. Specifically, I estimate how strongly the emission from Earth's nightside -- which is primarily caused by high-pressure sodium lamps emitting in the optical -- would be detectable in direct-imaging observations using the proposed LUVOIR and HabEx observatory architectures.

\section{The Emission Characteristics of City Lights on Earth}\label{section:data}

To begin, I first investigated the emission characteristics of city lights on Earth's nightside -- both the overall flux output and the spectrum of the emitted radiation at the top of the atmosphere. Recent space-based observations of Earth have generated several high-quality datasets of Earth's city lights as seen from low Earth orbit (Figure \ref{blackmarble}), which I used to make these estimates.

\subsection{The Overall Surface Flux of City Lights on Earth}

To determine the level of surface flux of the nightside city lights on Earth, I used data from the Visible Infrared Imaging Radiometer Suite (VIIRS) instrument on the Soumi National Polar-orbiting Partnership satellite. VIIRS is relatively unique, in that it possesses a Day/Night Band (DNB) which was specifically designed to make spatial and temporal measurements of nighttime lights on Earth. The DNB bandpass is relatively broad, and covers the optical from approximately 5000\,\AA\ to 9000\,\AA\ \citep{viirsdnbbandpass}. VIIRS DNB imagery has a resolution of 15 arcseconds, which corresponds to $0.742\times0.742$\, km pix$^{-1}$ on the Earth's surface at the spacecraft's nadir, and covers all longitudes from latitudes of -75$^\circ$ to +75$^\circ$.

\begin{figure*}[t]
\vskip 0.00in
\includegraphics[width=1.05\linewidth,clip]{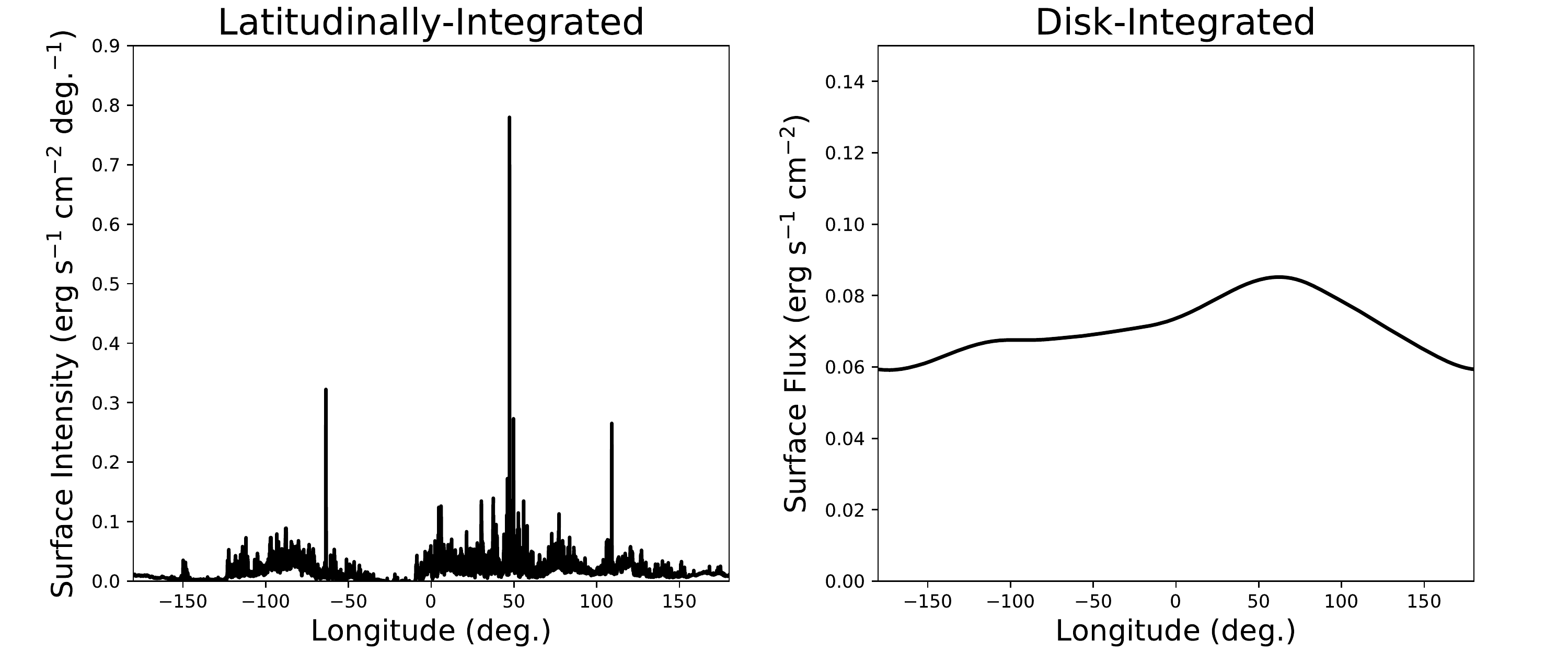}
\vskip -0.0in
\caption{While the latitudinally-integrated surface intensity from Earth (left panel) varies significantly as a function of longitude, the complete disk-integrated surface flux is relatively constant with longitude at $0.071$ erg s$^{-1}$ cm$^{-2}$. Note too that this integrated surface flux implies an globally-averaged surface intensity of about $0.02$ erg s$^{-1}$ cm$^{-2}$ sr$^{-1}$, which is significantly lower than the peak emission in city centers (e.g.., the few square kilometers around Times Square in New York City reach 47 erg s$^{-1}$ cm$^{-2}$ sr$^{-1}$). This reflects the fact that the total area covered by cities on Earth is low, at approximately 0.05\%.}
\label{latdisk}
\end{figure*}

One of the primary analysis hurdles in working with VIIRS DNB imagery is removing the effects of cloud cover, reflected light from the Moon, and airglow, so that the true flux from Earth's surface can be determined. The removal of these sources of additional light has been performed by the VIIRS team, who provide monthly and annual nighttime composite images without these background sources. Note that this also removes transient light sources from fires, lightning, and ships at sea. The typical flux precision in the annual DNB composites is $\pm0.03$ erg s$^{-1}$ cm$^{-2}$ sr$^{-1}$ pix$^{-1}$. As a practical illustration of the VIIRS DNB sensitivity, in a field experiment the instrument was able to successfully detect a 30 m$^{-2}$ tarp laid out on the ground in rural Puerto Rico, and illuminated from above by a single bright lamp \citep{blackmarbleATBD}.

I used the annual DNB composite from 2016, and numerically integrated the image to estimate both the latitudinally- and disk-integrated surface flux from the city lights on Earth (Figure \ref{latdisk}). One interesting thing to note in Figure \ref{latdisk} is that while the latitudinally-integrated surface flux varies significantly as a function of longitude, the total disk-integrated surface flux does not. For the rest of this work I therefore assumed that the surface flux from city lights did not depend on the sub-observer longitude, but rather was a constant value for all viewing geometries. For Earth, this is $0.071$ erg s$^{-1}$ cm$^{-2}$. Note, though, that both this average surface flux value and Figure \ref{latdisk} are for a perfectly cloud-free Earth. The typical cloud cover on Earth \citep{EarthCloudCover} will reduce this emission by approximately one-half, to $\mathrm{F_\oplus}=0.035$ erg s$^{-1}$ cm$^{-2}$.

In addition to clouds, the overall surface flux from nightside city lights will also change depending on the phase angle of the planet, and hence the fraction of the planetary nightside that is visible. Under the simplifying assumption that the disk-integrated surface flux of city lights is constant as a function of longitude (right panel of Figure \ref{latdisk}), then the overall surface flux from city lights on the nightside will scale as
\begin{equation}\label{eq:phaseangle}
\mathrm{F_p}(\theta) = \mathrm{F_{night}}\left(\frac{1+\cos(\theta)}{2}\right),
\end{equation}
where $mathrm{F_{night}}$ is the disk-integrated surface flux from the nightside of the planet (whatever the level of urbanization), and $\theta$ is the sub-observer longitude with $\theta=0^\circ$ defined to be exactly above the nightside hemisphere of the planet. This comes from integrating the $\cos\lambda$ longitude term in the disk-integration process, and the cosine dependence means that relatively high values for $\theta$ still give substantial emission fractions. For example, $\mathrm{F_p}(30^\circ)=0.93\,\mathrm{F_{night}}$, while $\mathrm{F_p}(60^\circ)=0.75\,\mathrm{F_{night}}$ and unsurprisingly $\mathrm{F_p}(90^\circ)=0.50\,\mathrm{F_{night}}$. Significant emission from nightside city lights is therefore possible to observe even if we see the planetary nightside from a substantial angle.

\subsubsection{How Surface Flux Changes with the Urbanization Fraction}

The total surface flux from city lights on Earth is relatively low, primarily because the total area covered by cities on Earth is also low. For example, the peak emission from New York City is 47 erg s$^{-1}$ cm$^{-2}$ sr$^{-1}$ in the few square kilometers around Times Square, while the peak emission from Tokyo is 35 erg s$^{-1}$ cm$^{-2}$ sr$^{-1}$ in the Shinjuku area. Based on the DNB composite images, only 0.05\% of Earth has a surface intensity that is on the order of these city cores, and I therefore considered that only 0.05\% of Earth's surface can be considered ``heavily'' urbanized. Due to this low fraction of the surface that is urbanized, though cities themselves are relatively bright overall the Earth does not show much emission from city lights.

However, it may be that advanced civilizations on exoplanets have built cities over significantly more of their planets' surface. These more urbanized planets would have a higher nightside brightness from city lights, and be correspondingly easier to detect. I therefore also considered how the urbanization fraction of a planet alters the overall surface flux from a planets' city lights, by applying a simple linear scaling based on the values I measured for Earth:
\begin{equation}\label{eq:urbanfrac}
\mathrm{F_p}(\mathrm{U}) = \left(\frac{\mathrm{F_{ecum}}-\mathrm{F_\oplus}}{1-\mathrm{U_\oplus}}\right)(\mathrm{U}-1)+\mathrm{F_{ecum}},
\end{equation}
where $\mathrm{U}$ is the urbanization fraction of the planet, $\mathrm{U_\oplus}$ is the urbanization fraction of Earth ($\mathrm{U_\oplus}=0.0005$, from above), and $\mathrm{F_{ecum}}$ is the surface flux of an ecumenopolis.

\subsubsection{Ecumenopoleis: Planet-wide Cities}

An ecumenopolis\footnote{Plural: ecumenopoleis}, or planet-wide city, is the limiting case where an entire planet is completely covered by a single massive city. For this I assumed that the entire surface of ecumenopoleis had a constant surface intensity similar to city centers on Earth, of $40$ erg s$^{-1}$ cm$^{-2}$ sr$^{-1}$. Accounting for typical cloud cover on Earth, this gave a surface flux for ecumenopoleis of $\mathrm{F_{ecum}}=62.8$ erg s$^{-1}$ cm$^{-2}$. Needless to say, such a planet would also have surface emission properties very different from the Earth \citep[e.g.,][]{schneider2018} -- particularly since both land and oceans would be covered by cities -- but for the purposes of this analysis I have neglected these effects and focus only on the increase in emission from city lights.

\begin{figure}[t]
\vskip 0.00in
\includegraphics[width=1.05\linewidth,clip]{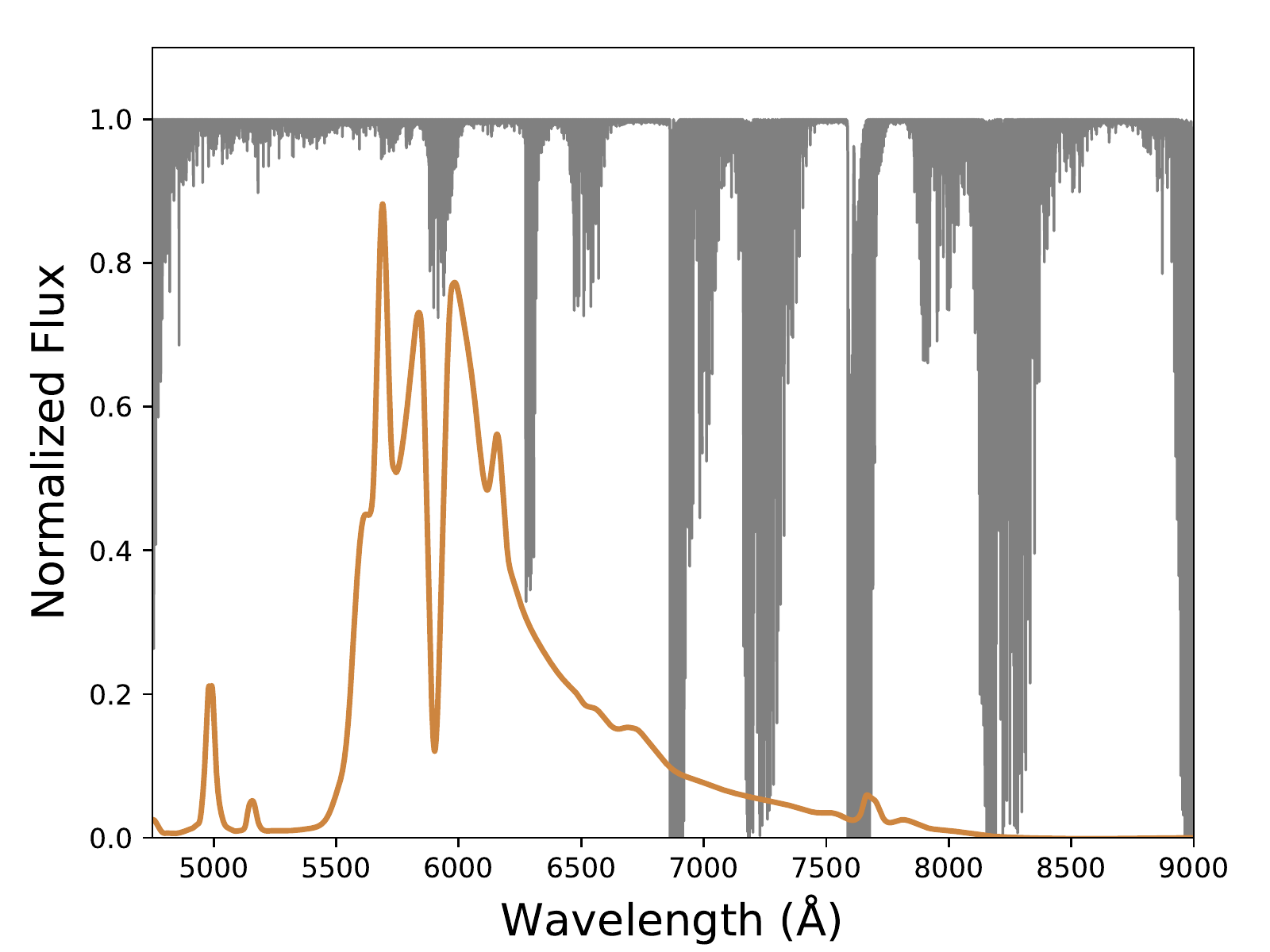}
\vskip -0.0in
\caption{The normalized emission spectrum of a typical high-pressure sodium street lamp (orange) shows large spectral signatures in the VIIRS DNB bandpass (5000\,\AA\ to 9000\,\AA). Also shown in grey is typical atmospheric transmission, which will obscure part of this emission when seen from space.}
\label{transpsec}
\end{figure}

\subsection{The Emission Spectrum of Earth's City Lights}

The dominant light source on Earth's night side is emission from street lights (or other area illumination lights), which reflect off nearby concrete and asphalt as seen from above. Modern street lights almost universally use high-pressure sodium (HPS) lamps of several hundred Watts, and though these are gradually being replaced by more energy efficient LEDs, for the purposes of this work I made the simplifying assumption that all of Earth's street lights are HPS lamps. I took as a typical HPS lamp emission spectrum that of the Sylvania 600W SHP-TS Super HPS bulb, which is provided on the product information page of Sylvania's website, and is also normalized and plotted here, in orange, in Figure \ref{transpsec}. Note that for this (and any other) bulb, while the input electrical power is 600W, the power emitted as light is significantly lower: approximately 130W in this case.

\begin{table*}[b]
\caption{Modeled Observatory Architectures and Properties}
\begin{tabular}{rcccc}
 & LUVOIR A & LUVOIR B & HabEx & HabEx w/SS \\
\hline
Diameter (m)      & 15                  & 8                   & 4                   & 4 \\
IWA               & 3.5 $\lambda$/D     & 3.5 $\lambda$/D     & 2.4 $\lambda$/D     & 60 mas \\
Raw Contrast      & $1.0\times10^{-10}$ & $1.0\times10^{-10}$ & $1.0\times10^{-10}$ & $1.0\times10^{-10}$\\
Avg. Throughput\textsuperscript{$\dag$}      & 0.3 & 0.4 & 0.4 & 0.7\\
Read Noise (e$^-$/pix) & 0.0 & 0.0 & 0.008 & 0.008\\  
Dark Current (e$^-$/pix/s) & 30\,$\mu$ & 30\,$\mu$ & 30\,$\mu$ & 30\,$\mu$\\
\hline
\multicolumn{5}{l}{\textsuperscript{$\dag$}\footnotesize{These average values are for reference only: the actual simulations included models of}}\\
\multicolumn{5}{l}{\footnotesize{coronagraphic throughput vs. separation that were based on the final study reports.}}
\end{tabular}
\vskip -0.1in
\label{obsprops}
\end{table*}

The emission spectrum from city lights on Earth as observed from space is also affected by the albedo spectrum of concrete, and the transmissivity of the atmosphere. For the former, most of the emission we would see from an exoplanet, and that we do see in the VIIRS DNB data, is caused by the light emitted by downward facing artificial lights reflecting back up, off nearby surfaces. On the Earth, these surfaces are predominately concrete buildings and roads, and I assumed that this will also be the case for exoplanet observations. That being said, the reflectance spectrum of concrete is effectively flat over the wavelengths of interest here \citep{lee2012}, so I did not include its effect in these calculations. The transmission of the atmosphere does have a significant effect on the emission spectrum, however, and I included a normalized model of optical transmissivity from \textsc{hitran} \citep[][shown in grey in Figure \ref{transpsec}]{hitran2016}.

\section{Observatory and Planetary Modeling}

I chose to model two proposed observatories to evaluate the detectability of city lights: the Large UV/Optical/IR Surveyor (LUVOIR) and the Habitable Exoplanet Observatory (HabEx). Both are large diameter space-based telescopes that would have dedicated, high-contrast, coronagraphs capable of directly imaging habitable Earth-like planets around nearby stars in the optical. I also considered the performance of direct-imaging instruments on some of the upcoming extremely large ground-based telescopes, and the Coronagraph Instrument on the Roman Space Telescope, but none of these would be capable of significant detections. LUVOIR and HabEx are the only proposed observatories with the contrast and inner-working angle (IWA) necessary to detect city lights on exoplanets.

\begin{figure}[bh!]
\vskip 0.00in
\includegraphics[width=1.05\linewidth,clip]{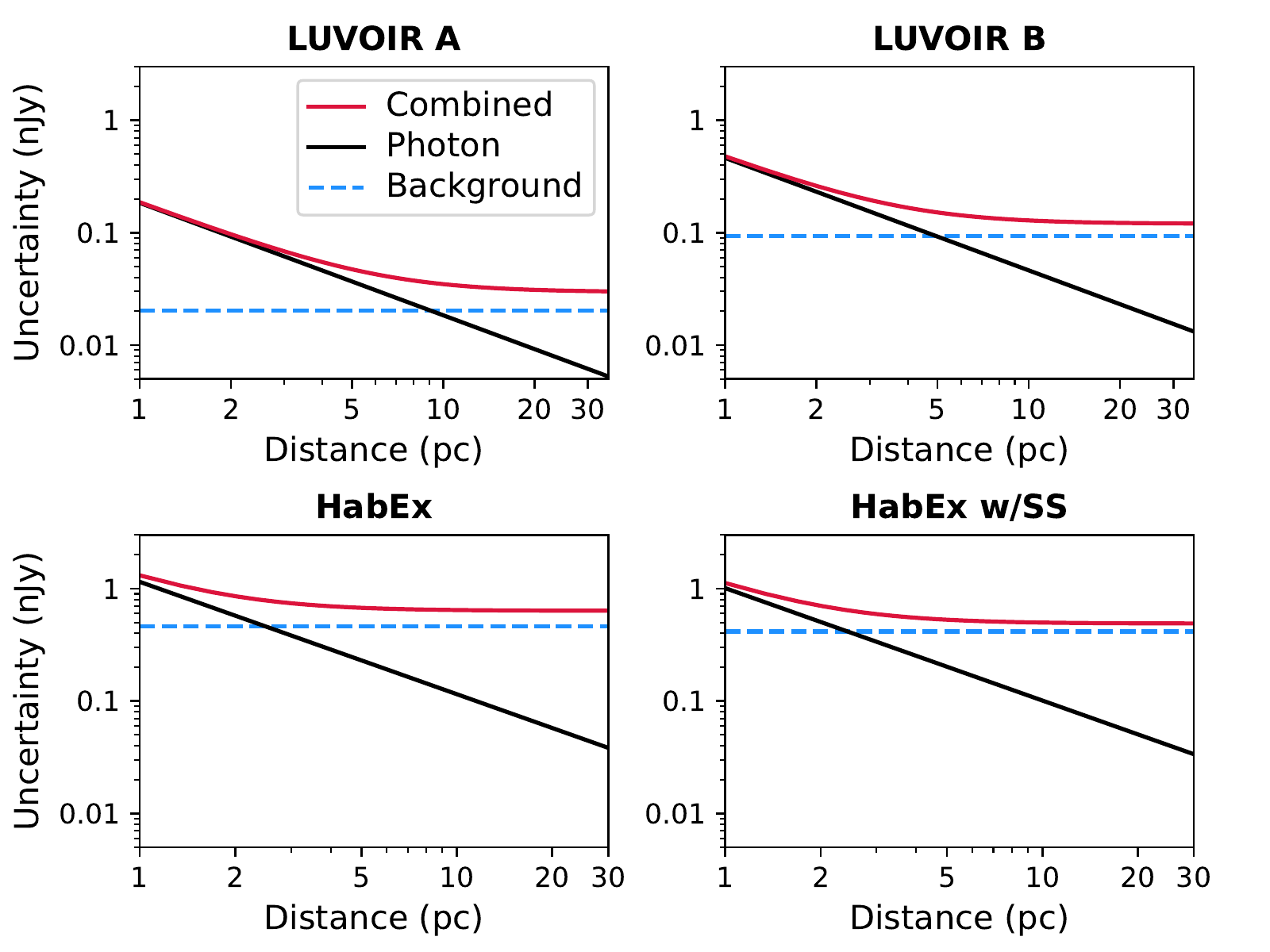}
\vskip -0.0in
\caption{The primary noise sources in these simulated 300 hours of direct-imaging observations are the photon noise from the combined planetary nightside and residual starlight, and the sky background. All four observatories become background-limited relatively quickly, though LUVOIR A and B both perform better than the HabEx architectures.}
\label{noise}
\end{figure}

I modeled two possible architectures for each observatory, based on the LUVOIR \citep{luvoirfinal} and HabEx \citep{habexfinal} final study reports. For LUVOIR, I considered the ECLIPS instrument on both LUVOIR A and LUVOIR B, and on HabEx I modeled the HCG instrument for both a baseline HabEx, and a HabEx with a star-shade. The properties I used for all four observatories are listed in Table 1. Note that the properties for LUVOIR A and B are the same, aside from the change in the primary mirror size, while the addition of a star-shade to HabEx only alters the IWA -- which becomes a constant 60 mas at all wavelengths. 

To calculate the expected performance of both sets of observatory architectures I used the Planetary Spectrum Generator \citep[PSG,][]{psg}, which includes both coronagraph and detector noise simulators. Specifically for both LUVOIR and HabEx, PSG includes models for how the coronagraphic throughput is expected to change as a function of the planet-star separation based on information in the final study reports. PSG also models the expected wavelength dependence of the coronagraphic throughput, so that in most cases (except for HabEx with a starshade) the measurement uncertainties increase towards the red. For these simulations -- which are used throughout the rest of the paper -- I assumed a baseline of 300 hours of observing time at a spectroscopic resolution of R=140. For reference, the LUVOIR Final Report expected 100 to 400 hours of observing time per target \citep{luvoirfinal}, while HabEx expected 200 to 800 hours per target \citep{habexfinal}. I also assumed an exozodiacal level of 3 zodis \citep{ertel2020}, and a Solar System zodiacal level of 1 zodi, in all cases.

Figure \ref{noise} illustrates the resulting noise models for an Earth-analog orbiting at 1 AU about a Sun-like star at various distances from the Solar System being spectroscopically observed at R=140 for 300 hours. Note that the units in Figure \ref{noise} are nJy, and these specific uncertainty values are for observations at 600\,nm. The two primary noise sources for these direct-imaging observations are the photon noise from the combined planetary nightside emission and residual unsuppressed starlight, and sky background -- with the detector noise being effectively zero. Given the low level of emission from planetary nightsides, these observations become background-limited relatively quickly, though as described in Section \ref{section:detect} it will be difficult for even LUVOIR A to significantly detect city lights beyond $\sim8$ pc.

I also used PSG to simulate the emission spectrum from an Earth-analog exoplanet without city lights. Specifically, I used the MERRA-2 atmospheric template \citep{merra2} available in PSG with ten sub-disk sampling regions, and I set the other planetary parameters to exactly match that of Earth. For simplicity, I did not consider possible atmospheric composition or temperature-profile changes caused by varying orbital distances, primary star type, planetary mass and radius, or possible tidal locking. 

\begin{figure*}[t]
\vskip 0.00in
\includegraphics[width=1\linewidth,clip]{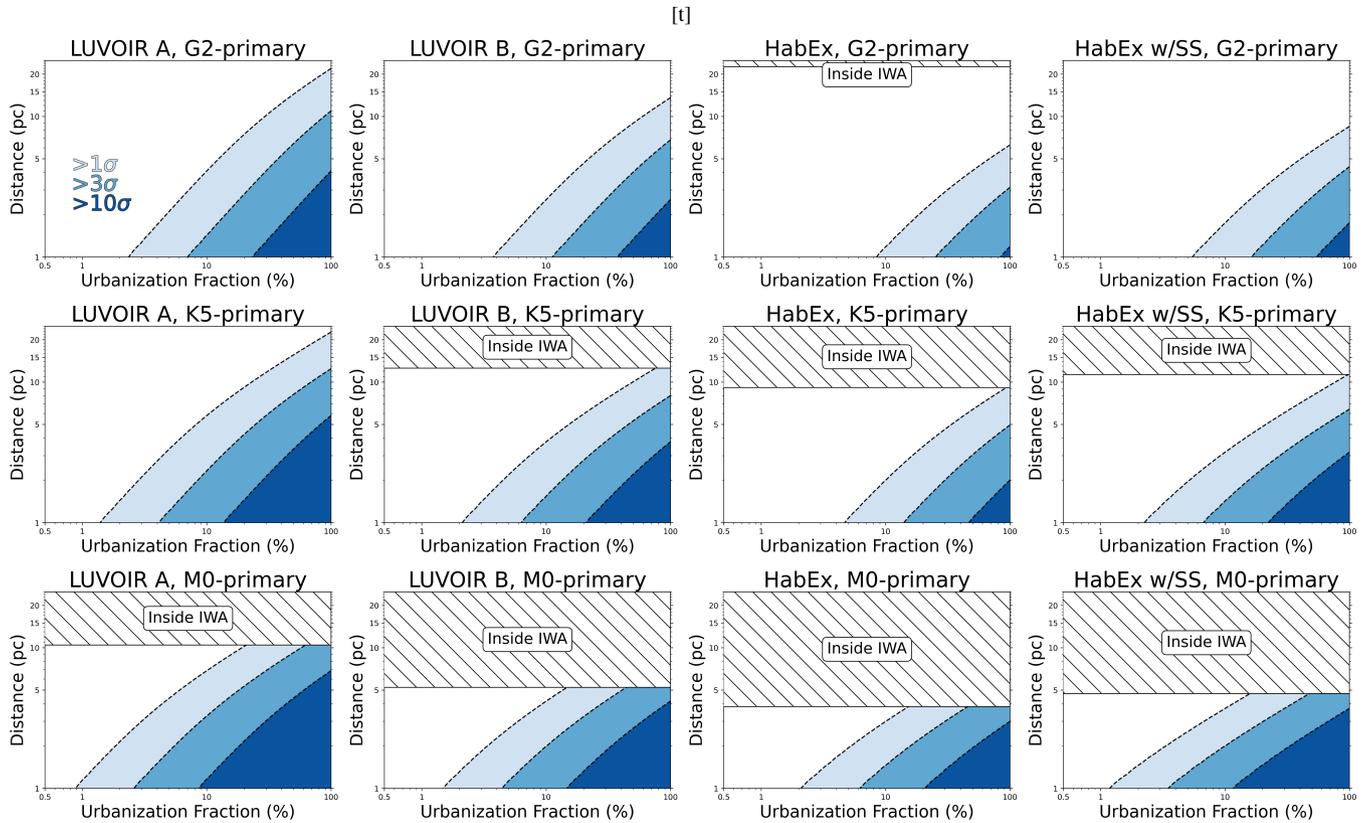}
\vskip -0.0in
\caption{The main trade-off in the detectability of city lights on a generic Earth-analog is between the spectral type of the parent star and the distance to which a planet would be imageable. Planets around later spectral types have higher planet-star contrast ratios which allow for city lights to be detected at lower urbanization fractions -- but the habitable zones around later stellar spectral types are also closer to the parent stars. This severely limits the distance at which city lights on planets orbiting M-dwarfs are detectable. The contours in these plots show the $1\,\sigma$ (light blue), $3\,\sigma$ (blue) and $10\,\sigma$ (dark blue) detect levels.}
\label{contours}
\end{figure*}

PSG also includes reflected star light in its simulated emission spectra, and I used the standard MERRA-2 albedo spectrum in all the simulations. For an Earth-analog orbiting a Sun-like star this causes the emission spectrum at the wavelengths of interest to be primarily leakage of reflected light from the dayside -- even at the nominal simulated phase angle of $\theta=30^\circ$ (recall that I defined $\theta=0^\circ$ to be directly above the nightside). The level of reflected light is strongly dependent on stellar type, and an Earth-analog orbiting an M-dwarf shows an emission spectrum at $\theta=30^\circ$ that is substantially fainter. For reference, the typical integrated surface flux at 600 nm for an Earth-analog around a Sun-like star at $\theta=30^\circ$ is 3 erg s$^{-1}$ cm$^{-2}$ \AA$^{-1}$, while a similar planet around Proxima Centuari has an integrated surface flux at 600 nm of only 0.02 erg s$^{-1}$ cm$^{-2}$ \AA$^{-1}$.

\section{Detecting City Lights}\label{section:detect}

I considered the detectability of nightside city lights in three ways. First, I investigated the general behavior of how detection significance changed as a function of stellar distance and planetary urbanization fraction for generic Earth-analogs around typical GKM-dwarfs. Second, I calculated the detectability of city lights for exoplanets orbiting stars within 8 pc of the Sun, assuming every star posses a generic Earth-analog. Third and finally, I estimated the detectability of city lights on two dozen known and potentially habitable terrestrial planets around stars close to the Sun (Table 2).

\begin{figure*}[t]
\vskip 0.00in
\includegraphics[width=1\linewidth,clip]{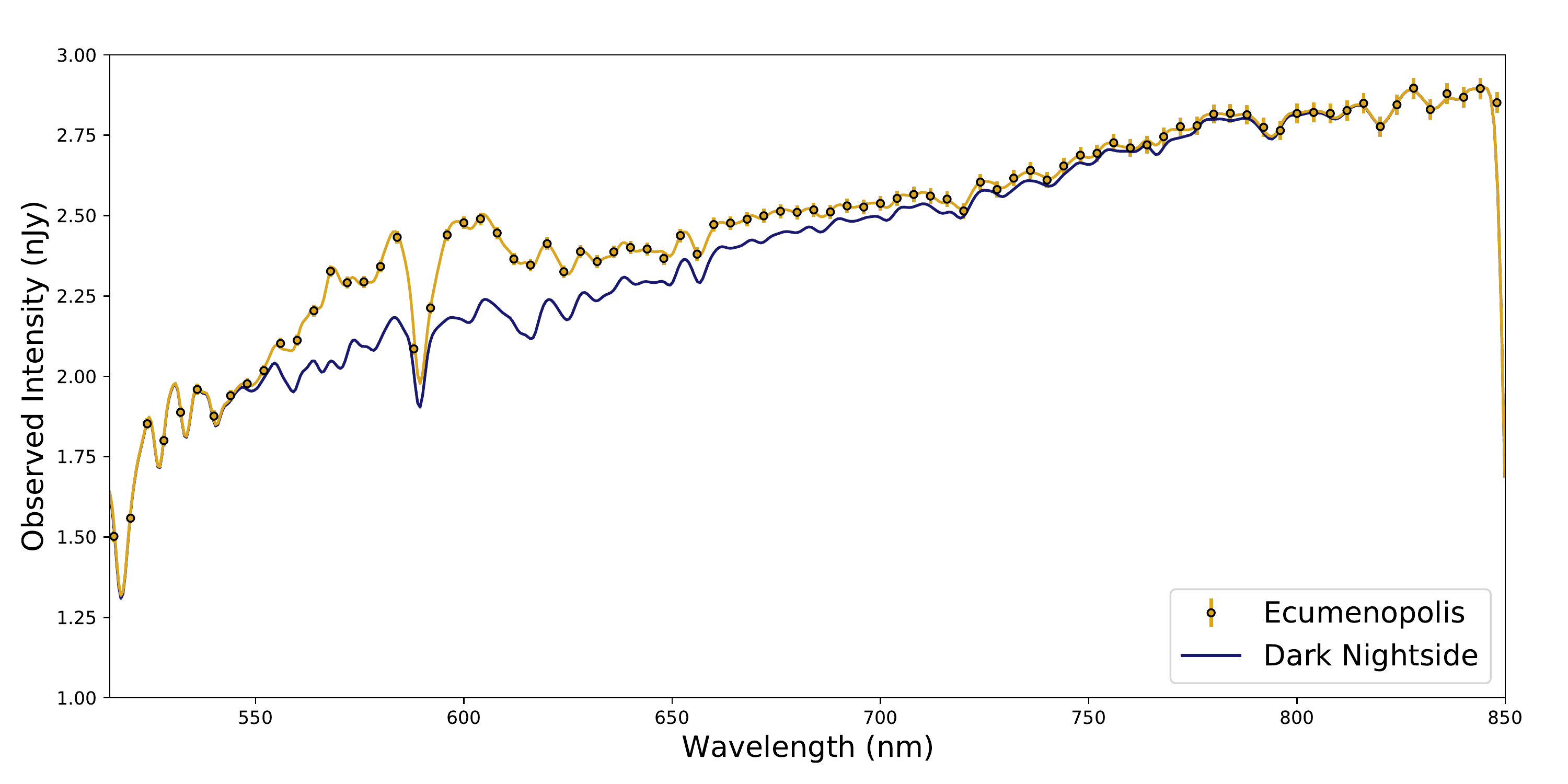}
\vskip -0.0in
\caption{An ecumenopolis -- a planet-wide city, see Section 2.1.2 -- in the habitable zone of Epsilon Indi would be easily detectable by LUVOIR A in 300 hours of observation, at $18\,\sigma$, as shown here. Note that almost all the planetary emission here is from the sliver of reflected daylight still visibile at the assumed phase angle of $\theta=30^\circ$. LUVOIR B would also strongly detect this planet ($19\,\sigma$), and it would likely be detectable by the baseline HabEx ($5\,\sigma$) and strongly detected by HabEx with a starshade ($11\,\sigma$). Generally, ecumenopoleis would be detectable out to considerable distances by all four observatory architectures. A full survey by LUVOIR A of 50 imageable stars within 8 pc would place a $1\,\sigma$ upper limit of $<1.9\%$ on the frequency of ecumenopolis planets in the Solar neighborhood -- assuming no detections. See the text of Section 4.1 for the upper limits using the other observatory architectures.}
\label{spec_ecumen}
\end{figure*}

In all these calculations I assumed that the city lights used were high-pressure sodium (HPS) lamps, similar to those used on Earth and as described in Section 2.2. HPS lamps emit primarily in the optical near 6000 \AA, and one might suppose that the inhabitants of a planet orbiting a significantly redder star than the Sun -- say Proxima b -- would choose redder lamps, with emission closer to the peak emission of their parent star. The reader should also note that LED lamps are beginning to replace HPS lamps as street lights on Earth, and the broad spectral emission of LEDs would likely make them more difficult to detect than HPS lights.

\subsection{Generic Earth-analogs Around Nearby Stars}

For the first and second cases, both of which assume a ``generic'' Earth-analog, I assumed the planets were perfect copies of Earth (i.e., mass, radius, atmosphere) orbiting at the outer edge of their stars' habitable zones. I took the outer edge of the habitable zone to be
\begin{equation}\label{eq:hadsep}
a_{\mathrm{hab}} = \left(\frac{L_*/L_\odot}{S_\mathrm{eff}}\right)^{0.5}\ \mathrm{AU},
\end{equation}
where $S_\mathrm{eff}$ is calculated based on Equation 4 in \cite{ravi2014} for the outer edge of the habitable zone. For the Sun, Equation \ref{eq:hadsep} gives a distance of 1.6\,AU. Note that the effect of placing these generic planets at the outer edge of the habitable zone, rather than at a separation where they would receive the same insolation as Earth, is to increase the distance at which an Earth-analog is detectable due to more planets being outside the IWAs of the observatories. The signal-to-noise ratio of the planets that are detected remains nearly constant at both locations.

I placed the generic planets at an orbital phase of $30^\circ$, unless the resulting, projected, star-planet separation was smaller than the IWA of the observatory. In that case I increased the orbital phase up to a maximum of $90^\circ$ or 1.2\,$\times$\,IWA -- whichever came first. I then scaled each planet's city light emission according to Equation \ref{eq:phaseangle} to account for the phase angle of the simulated observation. Most of the generic planets did end up at close to $30^\circ$, which meant that they emitted 0.98 of their total nightside flux towards Earth.

Note that the choice of a phase angle of $30^\circ$ does not greatly limit the number of planets that could potentially be imaged in this geometry. Recall that a phase angle of $30^\circ$ is possible for all planets with orbital inclinations of $60^\circ<i<90^\circ$. If we assume random inclinations for the newly discovered planets in these direct imaging surveys, then there is an 86.6\% chance that their orbital inclination will fall within this range \citep{beatty2010}. 

The main trade-off in the detectability of city lights on a generic Earth-analog is between the spectral type of the parent star and the distance to which a planet would be imageable (Figure \ref{contours}). Planets around later spectral types have higher planet-star contrast ratios which allow for city lights to be detected at lower urbanization fractions -- but the habitable zones around later stellar spectral types are also closer to the parent stars. This severely limits the distance at which city lights on planets orbiting M-dwarfs are detectable, as beyond 10 pc the apparent size of the habitable zone has shrunk to be inside the IWA of even LUVOIR A. 

Figures \ref{contours} and \ref{spec_ecumen} also illustrate how an ecumenopolis would be detectable around Sun-like stars by all four observatory architectures out to roughly 8 pc. Including later spectral types with closer limiting distances, there are 50 main-sequence stars within 8 pc that could host an ecumenopolis that would be detectable by LUVOIR A, 39 that would be detectable by LUVOIR B, 15 detectable by HabEx, and 32 detectable by HabEx with a starshade (Table \ref{tab:nearbystars}). A general survey of all these stars would be able to place meaningful upper limits on the frequency of ecumenopoleis in the Solar neighborhood, assuming no detections. A survey conducted using LUVOIR A would place $1\,\sigma$ and $3\,\sigma$ upper limits of $<2.2\%$ and $<10.7\%$ on the frequency of ecumenopoleis, while LUVOIR B would be able to place upper limits of $<2.8\%$ and $<13.4\%$, Habex would place upper limits of $<6.8\%$ and $<30.0\%$, and HabEx with a starshade would provide $1\,\sigma$ and $3\,\sigma$ upper limits of $<3.4\%$ and $<15.6\%$. This would provide valuable input into the Drake equation and the overall prevalence of extraterrestrial intelligence.

Tables \ref{tab:nearbystars} and \ref{tab:nearbystarsfracs} list the closest 50 stars around which a generic Earth-analog would show detectable emission from city lights to at least one of the observatory architectures at at least $3\,\sigma$. Note that several of the listed stars are in multiple star systems (e.g., Alpha Centauri A and B), but for simplicity I did not consider possible additional noise caused by the presence of other stars in the imaging field-of-view. Nearby M-dwarfs offer the best opportunities for detecting city lights from planets with low urbanization fractions that are similar to that of Earth. As shown in Table \ref{tab:nearbystarsfracs}, some of the best nearby targets are Proxima Centauri, Barnard's Star, Lalande 21185, Lacaille 9352 and Groombridge 34. All of these would show detectable signals from city lights at urbanization fractions of 0.005 to 0.030 in LUVOIR A imaging -- roughly 10 to 60 times higher than present-day Earth. 

Good targets for detecting city lights on planets around Sun-like stars are Alpha Centauri B, Epsilon Eridani, Tau Ceti, and Epsilon Indi A (Table \ref{tab:nearbystarsfracs}). All would be detectable at relatively low urbanization fractions by LUVOIR A ($\lesssim 5\%$ to $30\%$), LUVOIR B ($\lesssim 6\%$ to $25\%$), HabEx ($\lesssim 16\%$ to $70\%$), and HabEx with a starshade ($\lesssim 8\%$ to $34\%$). As with the nearby M-dwarfs, the two LUVOIR architectures or HabEx with a starshade are the three most capable observatory architectures. However, an ecumenopolis around any of these stars would be detectable by any of the observatories (e.g., Figure \ref{spec_ecumen}).

\subsection{Known Potentially Habitable Planets}

The above analysis concerned itself with generic Earth-analogs around nearby stars, since for many of these targets radial velocity and transit monitoring have not been able to firmly detect, or rule out, the presence of a potentially habitable terrestrial planet. However, we do know of about a dozen potentially habitable terrestrial planets around stars within 8 pc of the Sun, and I used the known properties of these planets to make specific predictions for the detectability of city lights on their surfaces. As with the generic Earth-analogs, I placed each of these known planets at a phase angle of $30^\circ$ -- since none have measured orbital inclinations -- and again increased this up to a maximum of $90^\circ$ or 1.2\,$\times$\,IWA to ensure it would be possible to image the planets.

To make this list, I took a relatively expansive view of what constitutes a ``potentially habitable'' planet. Specifically, I required $m_p\,\sin(i)<5$\,\mearth\ and that a planet's zero albedo, no redistribution, blackbody equilibrium temperature lay between $200\,\mathrm{K}<\mathrm{T}_{\mathrm{eq}}<400\,\mathrm{K}$. Given this wide range of parameter space, it is very likely that the atmospheric properties of terrestrial planets will vary significantly across this range -- but for simplicity I assumed that all of these planets had an Earth-like atmosphere and an Earth-like pressure-temperature profile. Making this assumption does not significantly affect the detectability estimates for city lights: the main effect of varying the atmospheric and planetary parameters would be to change the average cloud cover on the planet and hence the overall level of visible emission from city lights. However, cloud formation processes for Earth-like planets across this range of parameter space is currently not well understood.

\begin{figure*}[t]
\vskip 0.00in
\includegraphics[width=1\linewidth,clip]{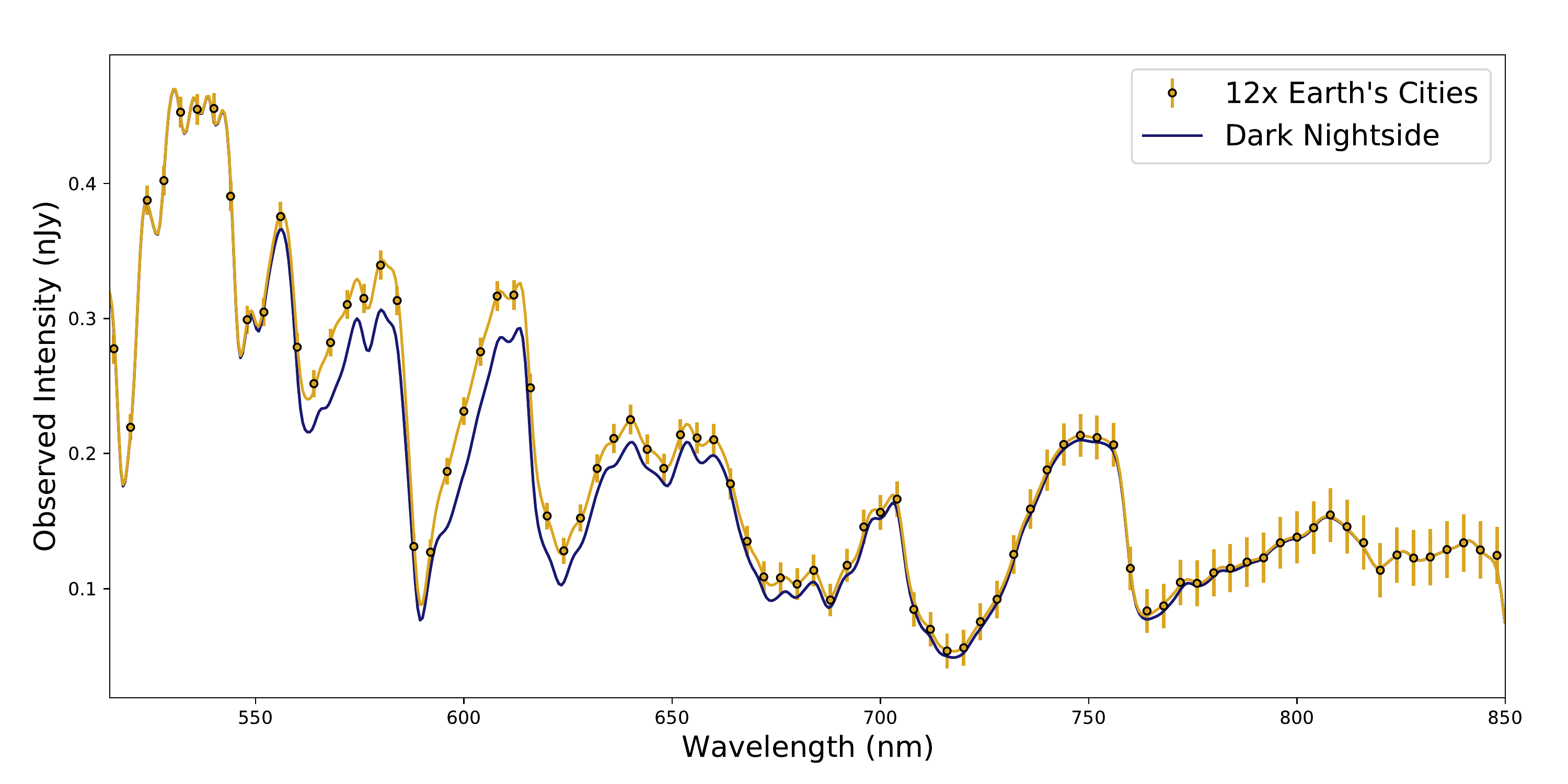}
\vskip -0.0in
\caption{Proxima b \citep{proxima} would be the best target to search for city lights using LUVOIR A. These simulated results illustrate the performance of 300 hours of LUVOIR A time at a spectroscopic resolution of R=140. The nightside of Proxima b would show detectable emission from city lights in LUVOIR A imaging (gold line and gold points) for urbanization fractions of 0.6\% and up. This is about twelve times the urbanization fraction on Earth, and may occur on Earth sometime in the mid-22nd-century \citep{landuse}. None of the other observatory architectures (LUVOIR B, or either HabEx) would be capable of imaging Proxima b, due to the planet being inside the IWA of those observatories.}
\label{spec_prox}
\end{figure*}

Since there are minimum mass measurements for all of these planets, I estimated their radii to better model the level of emission observable from Earth. To do so, I inverted the radius-mass relationship in \cite{weiss2014} for planets with radii less than 1.5\rearth, which corresponds to $m_p\lesssim5$\,\mearth, and then used the measured minimum masses as the true masses of the planets in the resulting calculations. 

\begin{table*}
\caption{Urbanization Fraction Detectable at $3\,\sigma$ in 300 Hours for Known Potentially Habitable Planets}
\begin{tabular}{lccccccc}
Planet & Mass (\mearth)  & $\mathrm{T}_{\mathrm{eq}}$ (K) & LUVOIR A & LUVOIR B & HabEx & HabEx w/SS\\
\hline
Proxima b       & $\geq1.17$    & 222 & 0.006 & -- & -- & --\\
Lalande 21185 b & $\geq2.69$    & 370 & 0.034 & -- & -- & --\\
Tau Ceti e      & $\geq3.93$    & 347 & 0.168 & 0.131 & 0.378 & 0.223\\
Tau Ceti f      & $\geq3.93$    & 220 & 0.146 & 0.096 & 0.297 & 0.147\\
Luyten's Star b & $2.89\pm0.26$ & 293 & 0.029 & 0.109 & 0.276 & 0.118\\
Kapteyn's Star b\textsuperscript{$\ddag$} & $\geq4.80$ & 335 & 0.042 & -- & -- & --\\
82 Eridani e    & $\geq4.77$    & 344 & 0.191 & 0.265 & 0.925 & 0.493\\
\hline
\multicolumn{5}{l}{\textsuperscript{$\ddag$}\footnotesize{Disputed.}}\\
\multicolumn{7}{l}{\footnotesize{\textsc{note} -- Entries with a ``--'' denote systems where the planet is within the inner working angle of the}}\\
\multicolumn{7}{l}{\footnotesize{coronagraph, and thus not imageable.}}
\end{tabular}
\label{tab:actualfracs}
\end{table*}

Note that the existence of some of these planets is disputed, such as Kaypten's Star b \citep{robertson2015,guinan2016}. Rather than attempting to adjudicate these disputes, I included all these planets in my calculations, but I do note when a dispute is present.

As shown in Table \ref{tab:actualfracs}, the best possible target and observatory combination would be imaging Proxima b \citep{proxima} using LUVOIR A, which would be capable of detecting an urbanization fraction of 0.006, or 0.6\% (Figure \ref{spec_prox}. This is about twelve times the urbanization fraction on Earth. Based on current land use trends, this level of urbanization may occur on Earth sometime in the mid-22nd-century \citep{landuse}. An ecumenopolis on Proxima b would be almost trivial to detect using LUVOIR A, at a nominal detection threshold of $586\,\sigma$. Unfortunately, neither LUVOIR B, nor HabEx, with or without its starshade, have the IWA necessary to image Proxima b.

Other nearby, known, targets of interest are the planets around Lalande 21185 b \citep{lalande21185_1,lalande21185_2}, Luyten's Star b \citep{luytenstar}, and Kapyten's Star b \citep{kapytensstar}, all of which would show detectable city lights in LUVOIR A data at an urbanization fraction of 2.9\% to 4.2\% (Table \ref{tab:actualfracs}). Of these, Luyten's Star b is likely to be the best candidate, since it is a confirmed super-Earth with a mass of $2.89\pm0.26$\,\mearth\ and an equilibrium temperature of 293\,K. The other two planets are possibilities, but Lalande 21185 b is on the very inner edge of the habitable zone and relatively hot ($\mathrm{T}_{\mathrm{eq}}$=370\,K), while the existence of Kapyten's Star b is disputed \citep[see above,][]{robertson2015,guinan2016}. 

The two outer planets orbiting Tau Ceti \citep[planets e and f,][]{tauceti_1,tauceti_2} are the best potentially habitable terrestrial targets around a Sun-like star, with minimum detectable urbanization fractions of 16.8\% and 14.6\%. However, both planets are relatively massive, at $m_p\,\sin(i)=4$\,\mearth \citep{tauceti_2}, and it is possible that neither are actually terrestrial planets.

As with the case of the generic Earth-analogs, extrapolating the values in Table \ref{tab:actualfracs} shows that if any of these known planets are an ecumenopolis they would be strongly detected by LUVOIR A. LUVOIR B and the two HabEx architectures, however, would have a more limited detection-space, since three of these known planets are inside the IWA of these observatories. The only detectable ecumenopoleis for LUVOIR B and either of the HabEx architectures would be the two Tau Ceti planets, Luyten's Star b, or 82 Eridani e.

\section{Summary and Conclusions}\label{section:conclusions}

As mentioned in the Introduction, the emission from nightside city lights on exoplanets is a compelling technosignature to search for using the proposed next-generation space telescopes LUVOIR and HabEx. The spectroscopic signature of Earth's city lights, as generated by high-pressure sodium lamps, is unique and difficult to cause via natural processes on a habitable terrestrial planet, and the idea of city lights requires little extrapolation from present-day Earth. This places the emission from nightside city lights high on the list of potential technosignatures to consider \citep{sheikh2020}. Generally speaking, the search for emission from city lights could be conducted by piggy-backing on searches for biosignatures with LUVOIR and HabEx, since the deep imaging required to detect either is roughly the same.

Here, I estimated the detectability of nightside city lights in the optical using 300 hours of direct imaging observations with the proposed architectures for the LUVOIR \cite{luvoirfinal} and HabEx \cite{habexfinal} observatories. For simplicity, I only modeled the detection of light from the kinds of high-pressure sodium lamps that are typically used as street lights on Earth, and I assumed a similar cloud covering fraction as Earth's. Though an Earth analog would not be detectable around even the closest stars, several nearby known potentially habitable exoplanets and nearby stars would show emission detectable at $3\,\sigma$ for relatively low urbanization fractions. Specifically:
\begin{enumerate}
    \item LUVOIR A would be capable of detecting city lights on Proxima b \citep{proxima} at an urbanization level of 0.6\%, or twelve times that of Earth (Figure \ref{spec_prox}), at $3\,\sigma$ confidence. Earth is expected to reach this urbanization level in around the mid-22nd-century.
    \item Other promising known exoplanets are Lalande 21185 b \citep{lalande21185_1,lalande21185_2}, Luyten's Star b \citep{luytenstar}, and Tau Ceti e and f \citep{tauceti_1,tauceti_2}. All would show detectable emission from city lights at urbanization levels of 3\% to 17\% in LUVOIR A imaging (Table \ref{tab:actualfracs}).
    \item Generically, city lights on Earth-analogs will be easiest to detect around nearby M-dwarfs (e.g., Proxima Centauri, Barnard's Star, or Lalande 21185), though the habitable zone around M-dwarfs quickly moves within the IWA of both LUVOIR and HabEx as distance increases (Figure \ref{contours}). Earth-analogs around Sun-like stars are imageable at larger distances, but the minimum detectable urbanization fraction is generally higher (Table \ref{tab:nearbystarsfracs}). Good Solar-like targets are Alpha Centauri, Epsilon Eridani, Tau Ceti, and Epsilon Indi A.
    \item An ecumenopolis, or planet-wide city, would be detectable by LUVOIR and HabEx out to a significant distance (Figures \ref{contours} and \ref{spec_ecumen}). A survey of the roughly 30 to 50 nearby stars within this distance (Table \ref{tab:nearbystars}) would be able to place a $1\,\sigma$ upper limit of $\lesssim2\%$ to $\lesssim4\%$, and a $3\,\sigma$ upper limit $\lesssim10\%$ to $\lesssim15\%$, on the frequency of ecumenopolis planets in the Solar neighborhood assuming no detections.
\end{enumerate}

The possibility of directly detecting technosignatures on the surfaces of potentially habitable exoplanets is thus starting to be in the realm of practicality. Perhaps unsurprisingly, the 15m LUVOIR A architecture would be the most capable observatory for detecting city lights on the nightsides of nearby exoplanets, though LUVOIR B or HabEx with a starshade would also have significantly sized detection spaces. Much of this proposed capability has been spurred by the goals of characterizing the atmospheres of and detecting biosignatures on potentially habitable exoplanets, but it also would afford us the opportunity to search for other, technological, signs of life. 

\section*{Acknowledgements}

The author wishes to thank Geronimo Villanueva and Ravi Kopparapu for helpful discussion during the preparation of this manuscript. This work has made use of the Planetary Spectrum Generator \citep{psg}, NASA's Astrophysics Data System, the Exoplanet Orbit Database and the Exoplanet Data Explorer at exoplanets.org \citep{exoplanetsorg}, the Extrasolar Planet Encyclopedia at exoplanet.eu \citep{exoplanetseu}, the SIMBAD database operated at CDS, Strasbourg, France \citep{simbad}, and the VizieR catalog access tool, CDS, Strasbourg, France \citep{vizier}.

\section*{Data Availability}

The data underlying this article will be shared on reasonable request to the corresponding author.

\newpage
\clearpage

\begin{table*}
\caption{Earth-analog Ecumenopolis Detection Significance (in $\sigma$) in 300 Hours}
\begin{tabular}{lccllll}
Star & Sp. Type  & Distance (pc) & LUVOIR A & LUVOIR B & HabEx & HabEx w/SS\\
\hline
Proxima Centauri\textsuperscript{$\dag$} &   M & 1.29 & 586.3 & -- & -- & --\\
Alpha Centauri A &   G & 1.35 & 29.2\textsuperscript{$*$} & 32.4 & 12.9 & 26.6\\
Alpha Centauri B &   K & 1.35 & 56.7 & 52.0 & 18.8 & 37.5\\
Barnard's Star &      M & 1.82 & 265.4 & 121.7 & 21.4 & 49.1\\
Lalande 21185\textsuperscript{$\dag$} &      M & 2.55 & 95.9 & 58.4 & 11.4 & 24.6\\
Ross 154 &           M & 2.97 & 84.3 & -- & -- & --\\
Epsilon Eridani &    K & 3.22 & 16.7 & 19.1 & 5.8 & 12.0\\
Lacaille 9352\textsuperscript{$\dag$} &      M & 3.29 & 51.6 & 35.6 & 6.9 & 15.1\\
Ross 128\textsuperscript{$\dag$} &           M & 3.34 & 62.4 & -- & -- & --\\
61 Cygni A &         K & 3.48 & 27.1 & 24.4 & 5.9 & 12.6\\
61 Cygni B &         K & 3.49 & 34.5 & 27.3 & 6.1 & 13.1\\
Struve 2398 A &      M & 3.53 & 50.8 & 32.8 & -- & 13.6\\
Struve 2398 B &      M & 3.53 & 52.0 & 32.8 & -- & 13.4\\
Groombridge 34 A &   M & 3.57 & 46.1 & 31.4 & 6.0 & 13.2\\
Groombridge 34 B &   M & 3.57 & 50.9 & 31.6 & -- & --\\
Tau Ceti\textsuperscript{$\dag$} &           G & 3.6 & 8.8 & 11.8 & 4.3 & 8.7\\
Epsilon Indi A &     K & 3.62 & 18.3 & 19.2 & 5.2 & 10.8\\
YZ Ceti\textsuperscript{$\dag$} &            M & 3.71 & 46.6 & -- & -- & --\\
Luyten's Star\textsuperscript{$\dag$} &       M & 3.8 & 43.6 & -- & -- & --\\
Kapteyn's Star\textsuperscript{$\dag$} &      M & 3.92 & 33.4 & 25.1 & 4.9 & 10.8\\
Lacaille 8760 &      M & 3.95 & 31.5 & 24.3 & 4.8 & 10.5\\
Kruger 60 A &        M & 4.03 & 37.1 & -- & -- & --\\
Wolf 1061\textsuperscript{$\dag$} &          M & 4.3 & 30.5 & -- & -- & --\\
WX Ursae Majoris &   M & 4.83 & 20.5 & 16.4 & 3.3 & 7.3\\
Groombridge 1618 &   K & 4.87 & 17.1 & 14.4 & 3.2 & 7.0\\
GJ 832 &             M & 4.97 & 19.6 & 15.5 & -- & 6.9\\
Omicron 2 Eridani &  K & 4.99 & 7.0 & 8.8 & 2.7 & 5.6\\
70 Ophiuchi A &      K & 5.12 & 6.3 & 8.1 & 2.5 & 5.3\\
70 Ophiuchi B &      K & 5.12 & 11.8 & 11.6 & 2.8 & 6.1\\
LHS 1723\textsuperscript{$\dag$} &           M & 5.37 & 16.3 & -- & -- & --\\
GJ 251\textsuperscript{$\dag$}  &            M & 5.58 & 14.7 & -- & -- & --\\
Sigma Draconis &     G & 5.76 & 5.1 & 6.5 & 2.1 & 4.4\\
GJ 229 A &           M & 5.76 & 12.9 & 10.9 & -- & 5.1\\
33 Librae A &        K & 5.86 & 7.7 & 8.1 & 2.1 & 4.6\\
Eta Cassiopeiae B &  K & 5.95 & 9.7 & 9.0 & 2.1 & 4.7\\
36 Ophiuchi C &      K & 5.98 & 7.4 & 7.8 & 2.1 & 4.5\\
HR 7703 A &          G & 6.02 & 5.4 & 6.5 & 2.0 & 4.2\\
82 Eridani\textsuperscript{$\dag$} &         G & 6.04 & 4.1 & 5.4 & 1.9 & 3.9\\
Delta Pavonis &      G & 6.1 & 3.4\textsuperscript{$*$} & 4.8 & 1.8 & 3.7\\
GJ 625 &             M & 6.47 & 9.7 & -- & -- & --\\
HR 8832 &            K & 6.55 & 5.8 & 6.3 & 1.7 & 3.7\\
Xi Bootis B &        K & 6.71 & 6.4 & 6.4 & 1.7 & 3.6\\
GJ 667 C &           M & 7.11 & 7.4 & -- & -- & --\\
GJ 105 A &           K & 7.18 & 4.4 & 4.9 & 1.4 & 3.1\\
107 Piscium &        K & 7.48 & 3.2 & 3.9 & 1.3 & 2.7\\
Fomalhaut &          F & 7.61 & 6.7 & 6.5 & 0.6 & 1.1\\
TW Piscis Austrini & K & 7.61 & 4.2 & 4.4 & 1.3 & 2.8\\
Vega &               F & 7.68 & 6.5 & 6.3 & 0.5 & 0.9\\
p Eridani &          K & 7.82 & 3.5 & 3.9 & 1.2 & 2.6\\
Tabit &              F & 8.07 & 5.6 & 5.6 & 0.9 & 1.8\\
\hline
\multicolumn{7}{l}{\textsuperscript{$*$}\footnotesize{The contrast ratio in these cases is below the systematic limit of $10^{-11}$ assumed by both the LUVOIR}}\\
\multicolumn{7}{l}{\footnotesize{and HabEx studies.}}\\
\multicolumn{7}{l}{\textsuperscript{$\dag$}\footnotesize{Has a known potentially habitable planet; see Table 2 for those systems which are imagable.}}\\
\multicolumn{7}{l}{\footnotesize{\textsc{note} -- Entries with a ``--'' denote systems where an Earth-analog would be within the inner working}}\\ 
\multicolumn{7}{l}{\footnotesize{angle of the coronagraph, and thus not imageable.}}\\
\end{tabular}
\label{tab:nearbystars}
\end{table*}

\begin{table*}
\caption{Urbanization Fraction (Earth=0.0005) Detectable at $3\,\sigma$ in 300 Hours}
\begin{tabular}{lccllll}
Star & Sp. Type  & Distance (pc) & LUVOIR A & LUVOIR B & HabEx & HabEx w/SS\\
\hline
Proxima Centauri\textsuperscript{$\dag$} &   M & 1.29 & 0.005 & -- & -- & --\\
Alpha Centauri A &   G & 1.35 & 0.103\textsuperscript{$*$} & 0.093\textsuperscript{$*$} & 0.232\textsuperscript{$*$} & 0.113\\
Alpha Centauri B &   K & 1.35 & 0.053 & 0.058 & 0.159 & 0.08\\
Barnard's Star &      M & 1.82 & 0.011 & 0.025 & 0.14 & 0.061\\
Lalande 21185\textsuperscript{$\dag$} &      M & 2.55 & 0.031 & 0.051 & 0.263 & 0.122\\
Ross 154 &           M & 2.97 & 0.036 & -- & -- & --\\
Epsilon Eridani &    K & 3.22 & 0.18 & 0.157 & 0.513 & 0.249\\
Lacaille 9352\textsuperscript{$\dag$} &      M & 3.29 & 0.058 & 0.084 & 0.432 & 0.199\\
Ross 128\textsuperscript{$\dag$} &           M & 3.34 & 0.048 & -- & -- & --\\
61 Cygni A &         K & 3.48 & 0.111 & 0.123 & 0.509 & 0.239\\
61 Cygni B &         K & 3.49 & 0.087 & 0.11 & 0.493 & 0.228\\
Struve 2398 A &      M & 3.53 & 0.059 & 0.091 & -- & 0.221\\
Struve 2398 B &      M & 3.53 & 0.058 & 0.092 & -- & 0.224\\
Groombridge 34 A &   M & 3.57 & 0.065 & 0.096 & 0.5 & 0.228\\
Groombridge 34 B &   M & 3.57 & 0.059 & 0.095 & -- & --\\
Tau Ceti\textsuperscript{$\dag$} &           G & 3.6 & 0.341 & 0.253\textsuperscript{$*$} & 0.705\textsuperscript{$*$} & 0.344\\
Epsilon Indi A &     K & 3.62 & 0.163 & 0.156 & 0.581 & 0.277\\
YZ Ceti\textsuperscript{$\dag$} &            M & 3.71 & 0.064 & -- & -- & --\\
Luyten's Star\textsuperscript{$\dag$} &       M & 3.8 & 0.069 & -- & -- & --\\
Kapteyn's Star\textsuperscript{$\dag$} &      M & 3.92 & 0.09 & 0.12 & 0.608 & 0.278\\
Lacaille 8760 &      M & 3.95 & 0.095 & 0.124 & 0.621 & 0.285\\
Kruger 60 A &        M & 4.03 & 0.081 & -- & -- & --\\
Wolf 1061\textsuperscript{$\dag$} &          M & 4.3 & 0.098 & -- & -- & --\\
WX Ursae Majoris &   M & 4.83 & 0.146 & 0.183 & 0.914 & 0.413\\
Groombridge 1618 &   K & 4.87 & 0.176 & 0.208 & 0.935 & 0.427\\
GJ 832\textsuperscript{$\dag$} &             M & 4.97 & 0.153 & 0.194 & -- & 0.436\\
Omicron 2 Eridani &  K & 4.99 & 0.431 & 0.342 & $>$1 & 0.531\\
70 Ophiuchi A &      K & 5.12 & 0.473 & 0.37 & $>$1 & 0.563\\
70 Ophiuchi B &      K & 5.12 & 0.254 & 0.259 & $>$1 & 0.49\\
LHS 1723\textsuperscript{$\dag$} &           M & 5.37 & 0.184 & -- & -- & --\\
GJ 251\textsuperscript{$\dag$}  &            M & 5.58 & 0.204 & -- & -- & --\\
Sigma Draconis &     G & 5.76 & 0.588 & 0.463 & $>$1 & 0.686\\
GJ 229 A &           M & 5.76 & 0.233 & 0.275 & -- & 0.587\\
33 Librae A &        K & 5.86 & 0.391 & 0.368 & $>$1 & 0.648\\
Eta Cassiopeiae B &  K & 5.95 & 0.309 & 0.335 & $>$1 & 0.636\\
36 Ophiuchi C &      K & 5.98 & 0.406 & 0.383 & $>$1 & 0.672\\
HR 7703 A &          G & 6.02 & 0.555 & 0.463 & $>$1 & 0.719\\
82 Eridani\textsuperscript{$\dag$} &         G & 6.04 & 0.738 & 0.555 & $>$1 & 0.77\\
Delta Pavonis &      G & 6.1 & 0.88\textsuperscript{$*$} & 0.629\textsuperscript{$*$} & $>$1\textsuperscript{$*$} & 0.815\\
GJ 625 &             M & 6.47 & 0.311 & -- & -- & --\\
HR 8832 &            K & 6.55 & 0.517 & 0.479 & $>$1 & 0.807\\
Xi Bootis B &        K & 6.71 & 0.47 & 0.469 & $>$1 & 0.822\\
GJ 667 C &           M & 7.11 & 0.404 & -- & -- & --\\
GJ 105 A &           K & 7.18 & 0.689 & 0.617 & $>$1 & 0.978\\
107 Piscium &        K & 7.48 & 0.932 & 0.763 & $>$1 & $>$1\\
Fomalhaut &          F & 7.61 & 0.45 & 0.463 & $>$1\textsuperscript{$*$} & $>$1\textsuperscript{$*$}\\
TW Piscis Austrini & K & 7.61 & 0.721 & 0.676 & $>$1 & $>$1\\
Vega &               F & 7.68 & 0.462 & 0.474 & $>$1\textsuperscript{$*$} & $>$1\textsuperscript{$*$}\\
p Eridani &          K & 7.82 & 0.869 & 0.771 & $>$1 & $>$1\\
Tabit &              F & 8.07 & 0.534 & 0.54 & $>$1\textsuperscript{$*$} & $>$1\textsuperscript{$*$}\\
\hline
\multicolumn{7}{l}{\textsuperscript{$*$}\footnotesize{The contrast ratio in these cases is below the systematic limit of $10^{-11}$ assumed by both the LUVOIR}}\\
\multicolumn{7}{l}{\footnotesize{and HabEx studies.}}\\
\multicolumn{7}{l}{\textsuperscript{$\dag$}\footnotesize{Has a known potentially habitable planet; see Table 2 for those systems which are imagable.}}\\
\multicolumn{7}{l}{\footnotesize{\textsc{note} -- Entries with a ``--'' denote systems where an Earth-analog would be within the inner working}}\\ 
\multicolumn{7}{l}{\footnotesize{angle of the coronagraph, and thus not imageable.}}\\
\end{tabular}
\label{tab:nearbystarsfracs}
\end{table*}

\newpage
\clearpage



\bibliographystyle{mnras}
\bibliography{references} 



\bsp	
\label{lastpage}
\end{document}